 \documentstyle[twocolumn,prc,aps,epsf,epsfig]{revtex}
 \begin{document}
 
\newcommand{\be}{\begin{eqnarray}}
\newcommand{\ee}{\end{eqnarray}}
\twocolumn[\hsize\textwidth\columnwidth\hsize\csname@twocolumnfalse\endcsname
\title{
Jet Quenching in High Energy Heavy Ion Collisions
\\ by  QCD Synchrotron-like Radiation
}

\author{  E.V. Shuryak and I. Zahed }
\address{
Department of Physics and Astronomy,\\
State University of New York, 
Stony Brook NY 11794-3800, USA
}

\date{\today}
\maketitle

\begin{abstract}
We consider synchrotron-like radiation in QCD by generalizing Schwinger's
treatment of quantum synchrotron radiation in QED to the case of a constant 
chromomagnetic field. We suggest a novel mechanism for {\em jet quenching} 
in heavy ion collisions, whereby high-$p_t$ partons get depleted  through 
strong (classical) color fields. The latters are encountered in the color
glass condensate or in the form of expanding shells of exploding sphalerons.
Unlike bremsstrahlung radiation through multiple soft rescattering, synchrotron
radiation converts a jet into a wide shower of soft gluons. 
We estimate the energy loss through this mechanism and suggest that it
contributes significantly to the unexpectedly strong jet quenching observed 
at RHIC.
\end{abstract}
\vspace{0.1in}
]
\newpage

\section{Introduction}
\label{sec_intro}
\subsection{Radiation in Various settings}

By synchrotron-like radiation we mean radiation emitted by a charge 
moving in an external field that is strong enough not to allow for
a perturbative treatment. The strong external field problem requires
the exact trajectories classically, and the exact propagators quantum
mechanically. Throughout, we will refer to synchrotron radiation as the
part of the radiation stemming from within the strong field region,
while the radiation from the outside field region (if present) will
be still referred to as bremsstrahlung radiation. For ultrarelativistic 
particles of energy $E=\gamma\,m$, bremsstrahlung radiation is usually
collimated along the particle trajectory in two cones 
of opening angle $1/\gamma\ll 1$, while synchrotron radiation diverges
away from the particle trajectory.

\begin{figure}[!h]
\epsfxsize=3.cm
\centerline{\epsffile{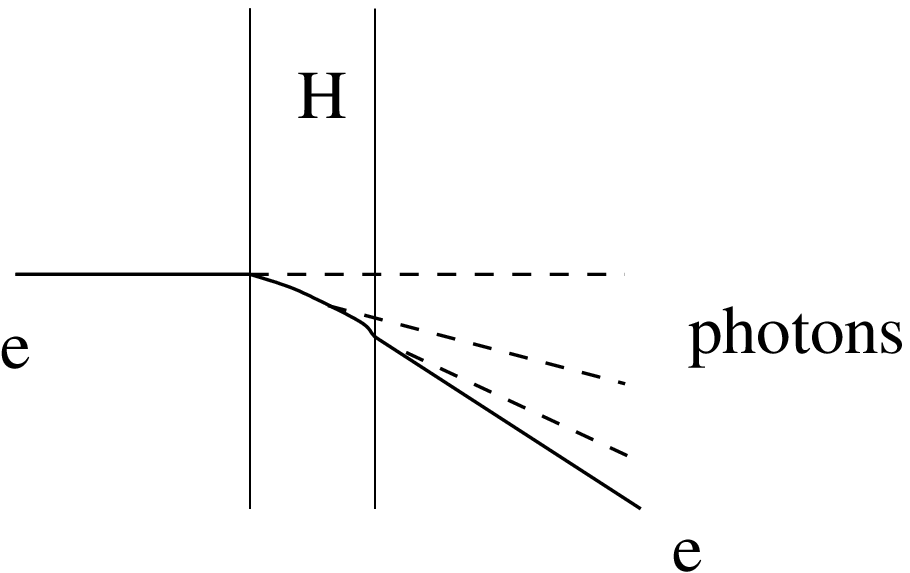}}
\epsfxsize=2.5cm
\centerline{\epsffile{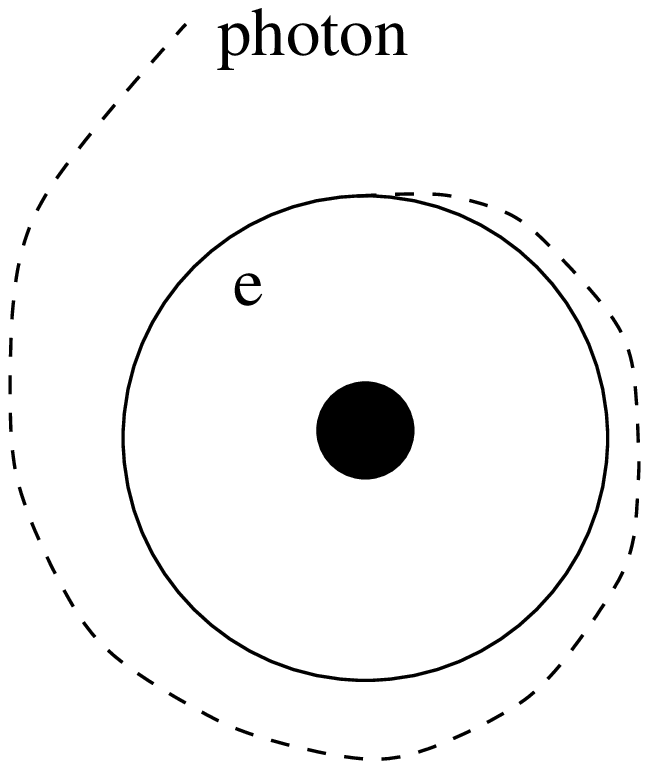}}
\epsfxsize=3.cm
\centerline{\epsffile{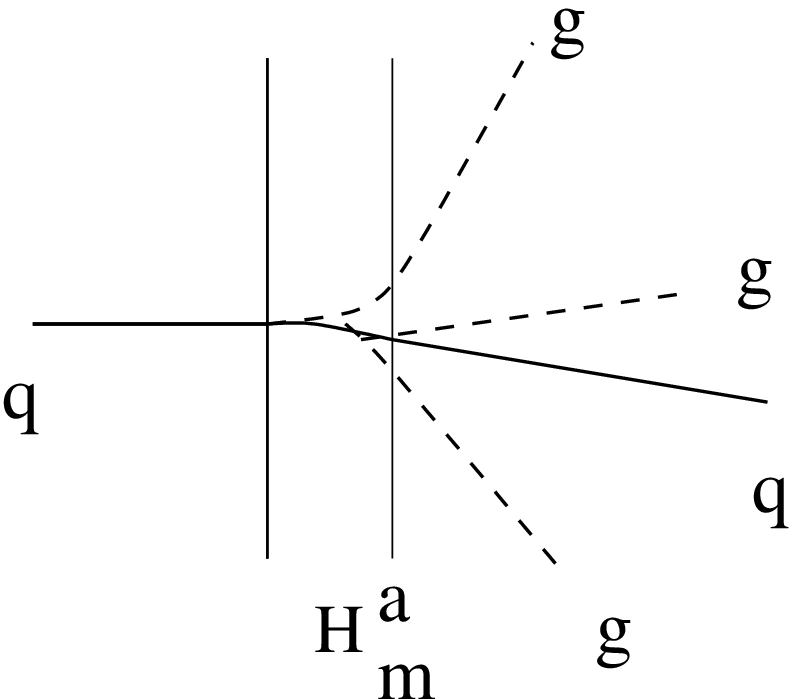}}
\vskip 0.05in
\caption[]{
 \label{fig_magnets}
Schematic representation of synchrotron-like radiation in three cases:
(a) in a magnetic field; (b) in a gravitational field (rotating charge around a
black hole); (c) in a chromomagnetic field.}
\end{figure}

In QED synchrotron radiation usually takes place within a magnet
as illustrated in Fig.~\ref{fig_magnets}a. The classical trajectory 
includes the circular bending between the incoming and outgoing straight
lines, and the radiation is emitted tangent to the arc length. Photons
move on straight lines since in QED they are not affected by the 
external magnetic field.

Another interesting case of classical synchrotron-like radiation 
is that of an ultrarelativistic rotating charge in a strong gravitational
field such as the one encountered near the horizon of a black-hole
\cite{Kh_Shu} (and subsequent literature). In this case, both the charge and
the photon are gravitationally deflected as illustrated in 
Fig.~\ref{fig_magnets}b. The result is a significant reduction of the
radiation loss: the total radiation yield is reduced by a factor of 
$\gamma^2$ in comparison to the yield from standard synchrotron
radiation for the same curvature. Also, the radiation length for
each particular direction is actually the entire circle, not just an
arc length of order $1/\gamma$ as in the magnet.

In this paper we consider an ultrarelativistic charge in QCD
(parton, quark, gluon) going through a constant chromomagnetic field
as illustrated in Fig.~\ref{fig_magnets}c.  The motion of the initial
charge and the ensuing radiation are both strongly affected by the 
chromomagnetic field. If classical geometrical optics can be used~\cite{Wong}
(and for hard jet quenching it is not the case) the classical motion of the
external particle and the radiation is described by non-trivial trajectories. 
For soft radiation, the emitted gluons are bent along circular paths of smaller
curvature radii. As a result, part of the radiation gets trapped in the near
field region and never makes it to infinity. This makes standard radiation
calculations obsolete. Moreover, the radiated gluons carry different charges
and therefore move in different directions. The QCD chromomagnetic field 
resolves both momentum and color thereby acting as a double (squared)
Newtonian prism.

Another instructive classical description to keep in mind is the
instantaneous distribution of the radiation field in the
ultrarelativistic case. In QED the nonzero field strength is mostly
located around a spiral-like curve known as the evolute of the circle. 
Since both the charge and the photon moves light-like, the distance
from the radiation point to the charge is the same as to the photon 
along a straight line. Details of the field distribution for QED synchrotron 
radiation can be found in~\cite{Tsien}. Its analog for QCD
in the simplest abelian-type external field when the radiated gluons 
undergo planar rotation has a cycloidal shape. The radiation field
is the same at any instant of time but moves (together with the charge)
at the speed of light.

\subsection{Synchrotron Radiation versus\\
 Bremsstrahlung Radiation in QED}
\label{sec_QED}

The number of soft prompt gluons per frequency $\omega$ expected from 
bremsstrahlung and synchrotron radiation can be estimated classically.
The Lienard expression for the power radiated from a dipole

\be 
P=-{2 e^2\over 3 m^2}({dp_\mu \over ds})^2 
\label{L1}
\ee
was derived in 1898. It holds even ultrarelativistically since the power
(energy over time) is relativistically invariant. The acceleration in 
(\ref{L1}) relates to the field strength $F$ and the particle energy $E$ 
by the Lorentz formula. Classically, the energy loss depends quadratically
on both $E$ and $F$, i.e. $P\sim e^4 F^2 E^2/m^4$. This result does not hold
at very large energies though. Indeed, it only holds in the range~\cite{LL}

\be \label{eq_condition}
 e^3 (F/m^2) (E/m) \ll 1  \,\,,
\ee
where we have separated 3 dimensionless physical factors: 
the coupling constant, the field strength in units of $m^2$ and
the relativistic gamma factor. For classical
QED synchrotron radiation in accelerators, this condition 
of course holds even with $\gamma=E/m\sim10^5$ (LEP).
The reason is the small electromagnetic coupling and the
small magnetic fields $H/m^2\ll 1$.

In the QCD problem we are interested in, the fields are strong,
the external charge is massless and the induced radiation recoil 
is large. The condition (\ref{eq_condition}) does not hold, and
we have to reassess the power radiated from first principles. 
However, we expect that qualitatively the differential spectra 
to follow general lore at small frequencies. In particular, the
number of bremsstrahlung photons per frequency should be of order

\be
\frac{dN_B}{d\omega}\sim &&\frac{\alpha}{\pi} \,\frac{1}{\omega} \ee 
while the number of synchrotron photon per frequency (per orbit) 
should be of order

\be
\frac{dN_S}{d\omega}\sim &&\frac{\alpha}{\pi} 
\,\frac{(\omega/\omega_0)^{1/3}}{\omega}\,\,,
\label{2}
\ee
where $\omega_0=eH/E$ is the synchrotron frequency with $E$ the
energy of the external charge. At very low frequencies bremsstrahlung 
of course dominates, but not at large frequencies.
Integrating (\ref{2}) up to some maximal frequency yields

\be
&&N_S\sim \frac{\alpha}{\pi}\, (\omega_{\rm max}/\omega_0)^{1/3}
\label{3}
\ee
which amounts to an energy loss
$\Delta E_S \sim \alpha \,\omega_{\rm
max}^{4/3}/\omega_0^{1/3}$. The maximum frequency is bounded
by the classical characteristic frequency 
$\omega_{\rm c}=3\gamma^3\omega_0$. As a result, the maximum 
number of synchrotron photons emitted per cycle is $N_S\sim \alpha \gamma$,
which is seen to grow linearly with the energy. The maximum synchrotron
energy loss $\Delta E_S\sim \alpha\, \gamma^4\, \omega_0$ grows cubicly with
the energy. In comparison, the power following from the Lienard
expression (\ref{L1}) grows quadratically with the energy 
$\Delta E\sim \alpha\, E^2$, and since one circle takes time
proportional to E, both expressions agree.

The above qualitative reasoning is entirely classical. In quantum 
mechanics the photon back reaction cannot be ignored, especially 
for large energy losses with $\omega \sim E$ (in units $c=\hbar=1$).
Although in modern accelerators like LEP $\gamma \sim 10^5$ resulting
into an enhancement in radiation per frequency that is about 15 orders
of magnitude high, the photon energies are usually much smaller than 
the energy of the rotating particle and the back reaction can be ignored.

In the QCD case to be considered below, this is not true. Using
quantum cutoff
$\omega_{\rm max} =E\ll \omega_c$, we then find that the number of
bremsstralung photons grows logarithmically with the energy 
$N_B\sim \alpha\, {\rm ln}\, (E/\omega_{\rm min})$ while the energy
loss grows linearly with the energy $\Delta E_B\sim \alpha\,E$. These
results are in total agreement with the ones obtained through standard
quantum calculations such as Feynman graphs. For synchrotron radiation
a cut at $\omega_{\rm max} =E$ yields 

\be 
N_S\sim \alpha\, (E^2/eH)^{1/3} 
\ee
and  the energy loss for a fixed width of the field region $\Delta z$
(rather than a circle)~\footnote{It is not small in 
comparison  to the radiation length: $\Delta z > m/eH$.}
 
\be 
\Delta E_S \sim e^2 (e H)^{2/3} \Delta z \, E^{2/3} \,\,,
\ee
which is seen to grow less than linearly. As a result, the
relative energy loss by synchrotron radiation $\Delta E_S/E\sim 1/E^{1/3}$
is found to decrease with energy, although with a small power.

The present qualitative 
estimates are in agreement with the full quantum calculations
to be described below. They show that bremsstrahlung radiation
dominates over synchrotron radiation at large energies, although
with a small relative power of $1/3$. However, we recall that 
subsequent bremsstrahlung radiations at small angle interfere
with each other and are further reduced by the Landau-Pomeranchuk-Migdal
(LPM) effect, while the synchrotron emitted quanta are lost once and
for all without further suppression.

\subsection{Classical fields in heavy ion collisions}

Synchrotron-like radiation may take place in QCD settings whenever
classical and strong non-abelian fields can be formed. We think 
that such strong semiclassical gauge fields with amplitudes $A\sim 1/g$,
naturally arise in the early stages of ultrarelativistic heavy-ion 
collisions. Currently, there are two first principle QCD reasonings
in favor of prompt and strong non-abelian gauge fields 
in heavy ion collisions.

The first reasoning by McLerran and Venugopalan~\cite{McLV}
suggests that the prompt phase of a heavy ion collision is a Color
Glass Condensate (CGC). Schematically, consider a small element 
$\Delta x_\perp$ of the 
transverse plane defined by the disk-shaped boosted heavy nuclei.
At high energy, the hadronic structure functions increase at small
parton momentum fraction $x=p/E\rightarrow 0$. This increase results 
in an increase in the density $n_Q$ of color charges in the transverse 
2-dimensional plane. The number of charges becomes ultimatly large
$N_Q=n_Q\,\Delta x_\perp\gg 1$, with a total charge of order ${\bf Q}\sim 
\sqrt{N_Q}$~\footnote{We are assuming that all partons originate from
different nucleons and add randomly.}. The large initial 
color charge ${\bf Q}$ is at the origin of the classical color field.
The smallest transverse size $\Delta x_\perp\sim 1/Q_s^2$ is fixed
by the parton saturation scale $Q_s$ reached when the gauge field
amplitude satisfies $\partial\, A \sim g\, A^2$~\cite{saturation}.
In this approach, the {\it virtual} classical fields are attached
to the nuclear wavefunction and become real after being stripped 
off by the heavy ion collision.

The second reasoning suggests rather that the heavy ion collision
converts the virtual classical vacuum fields to real fields. In other
words the classical fields originate from the wavefunctional of the
QCD vacuum. Yang-Mills fields have a rich topological structure in the
vacuum, with large amplitude tunneling described classically by
instantons~\cite{BPST} (for a review see~\cite{SS_98}). Unlike 
the virtual fields close to zero strength which can be made real
by perturbation theory, the virtual~\footnote{In the so called Landau
method used e.g. in \cite{JSZ}, those fields  even have additional
singularity, essential  for the evaluation of the tunneling
cross section.} tunneling fields deep under
the barrier can only be made real after an energy deposition
sufficient to jump $over$ the barrier.
As we have argued recently~\cite{OCS,Shu_how,JSZ,SZ_quarks} using
arguments in favor of the relevance of instantons to high energy 
hadron-hadron scattering in the semi-hard (soft pomeron) region
~\cite{SZ} and insights from electroweak 
theory~\cite{weakinst}, these fields are released in Minkowski
space in purely chromomagnetic form, the relatives of the 
{\it sphalerons} in the electroweak theory. Indeed, much like their
relatives they are classically unstable and evolve (explode) into
a thin shell of a coherent color field.

The QCD vacuum is filled with relatively small instantons (and also 
antiinstantons),
with an average size $\rho\sim 1/3$ fm which is small in comparison to their
relative separation which is of the order of 1 fm. The QCD vacuum is
characterized by the small dimensionless instanton diluteness parameter
$\kappa = n\rho^4 \sim 0.01$. Therefore and  immediatly upon
their release the QCD sphalerons are expected to be in a dilute phase as well.
When produced in bunch like in a heavy ion collision, the QCD sphalerons
evolve pretty much unshattered for a time of the order of 1 fm before they
collide and get destroyed. More details for this process will be given
below.

In a recent paper by one of us \cite{Shu_how} the idea of jet quenching
on coherent classical fields has been first discussed. The importance
of the synchrotron-like radiation is due to both of us and was briefly
advertised in~\cite{Shu_how}. The present paper elaborates further on 
this idea and presents detailed quantum calculations for jet quenching by
synchrotron-like radiation.

\section{QCD Synchrotron Radiation}

In this section we proceed to estimate the QCD synchrotron radiation
in a constant and Abelian-like chromomagnetic field. We will derive
the exact classical and first quantum correction in the regime 
$\omega/E<1$, and will provide an approximate expression for all
frequencies $\omega$.

The problem of quantum synchrotron radiation in QED was addressed
in a fundamental way by Schwinger~\cite{S72} using the mass operator
formalism. In this section, we extend this approach to the 
quantum synchrotron radiation in QCD. Two essential differences
between the QED and QCD problem: i. the non-Abelian nature of the
charge in QCD; ii. the emitted radiation also undergoes magnetic
deflection. A quantum calculation is required in strong chromomagnetic
fields owing to potentially large recoil corrections, essential for
large jet quenching. The power radiated will be sought through the
mass operator as

\be
-\frac{1}{E}\,{\rm Im}\,{\bf M}_{aa} =
\int\,\frac{d\omega}{\omega}\,{\bf P}_{aa} (w)\,\,\,
\label{SR8}
\ee
after pertinent kinematical identifications.

\subsection{Abelian Chromomagnetic Field}

For simplicity, we consider QCD synchrotron radiation in a constant
and homogeneous chromomagnetic field 

\be
G^a_{\mu\nu} (x)  = \delta^{a8}\,G_{\mu\nu} 
\label{SR1}
\ee
where the abelian field strength corresponds to a constant
magnetic field in the 3-direction, $G_{12}=-G_{21}=H$. The 
background gauge field associated to (\ref{SR1}) is

\be
A_\mu^a (x) = \delta^{a8}\,A_\mu (x) = \delta^{a8}\,\delta_{\mu 2}\,H\,x_1\,\,.
\label{SR2}
\ee
With our choice of the chromomagnetic background along the
8th color direction, the quarks and gluons can be diagonalized.
The diagonal quarks in the fundamental representation
carry color ($a=1,2,3$)

\be
e_a = g\,(T^8)_{aa} = \frac{g\sqrt{3}}{6}\,(1,1,-2)
\label{SR2x}
\ee
and the diagonal gluons in the ajoint representation 
carry color

\be
g_A= (-1)^A\,\frac{g\sqrt{3}}2\,\,,
\label{SRxx}
\ee
for $A=4,5,6,7$ and $g_A=0$ for $A=1,2,3,8$. 
These two cases, as will be shown below, lead to qualitatively different
radiation. The second case is basically QED-like.

Quantum synchrotron radiation will be sought for quarks and
gluons interacting to {\rm all} orders in $H$ but to leading
order in $\alpha=g^2/4\pi$ between the quantized fields. We
now present briefly the spin-1/2 case and discuss extensively
the spin-0 case. In the semiclassical limit, both  spins
radiate at the same rate.

\subsection{Spin $0, 1/2$ Jets}

Following Schwinger~\cite{S72}, to lowest order in perturbation 
theory the quark mass operator in the chromomagnetic field reads

\be
&&{\bf M}_{aa} =
ig^2\,(T^A)_{ab}\,(T^A)_{ba}\,\int\,\frac{d^4k}{(2\pi)^4}\,
\int_{0-i0}^{\infty-i0}\,\frac{ds}{{\rm cos}(g_A\,Hs)}\nonumber\\
&&\times\,
e^{-is(k^2-k_\perp^2({{\rm tan}(\,g_A Hs)}/({g_A Hs}) -1))}\,
\left(e^{2g_A sG}\right)_{\mu\nu}\,\,\Phi (g_A)\nonumber\\
&&\times\,\gamma^{\mu}\,(\gamma\cdot (\Pi_b-k)-m)^{-1}\gamma^{\nu}\,\,.
\label{SR3}
\ee
We have defined the quark 4-momentum operator as

\be
\Pi_{b\,\mu} (x)  = i\,\partial_\mu- e_b\,A_\mu (x)
\label{SR4}
\ee
and the Bohm-Aharanov  line

\be
\Phi (g_A; x,y) = e^{i\frac{g_A\,H}2\,(x_1+y_1)(x_2-y_2)}\,\,.
\label{SR5}
\ee
The Bohm-Aharanov phase enforces gauge-invariance in the
mass operator, but does not contribute to the radiation. Indeed,
for an initial color-a quark emitting a color-b quark plus a
color-A gluon,

\be
\Phi (e_a; x, y) = \Phi (e_b; x, y)\times \Phi (g_A ; x, y)
\label{SR6}
\ee
showing that the Bohm-Aharanov line in (\ref{SR3}) on the gluon,
can be redistributed to compensate the analogue ones on the quarks.
This procedure will be assumed throughout, and thereby the gluon
$\Phi$ contribution reshuffled.

The occurrence of the synchrotron poles
in the gluon propagator (\ref{SR3}) implies that the 
s-integration is infinitesimally shifted below the real axis in the 
complex s-plane. The prescription follows the causal prescription
for the free propagator, 

\be
\frac 1{k^2+i0}=i\,\int_{0-i0}^{\infty-i0}\,ds\,e^{-is\,k^2}\,\,.
\ee
Also, 
since $G$ is an antisymmetric matrix, its eigenvalues $\pm iH$
are complex. The color precession factor is

\be
e^{2g_AsG}\rightarrow e^{\pm is\,(2g_AH)}\,\,.
\label{SR5x}
\ee
There is a subtlety due to 
the positive sign in (\ref{SR5x}) for $k=0$, which is
the analogue of the tachyonic mode of a spin-1 coupled
to a constant chromomagnetic field in the first quantized
approach. This mode is at the origin of the well-known
Savvidy  instability in QCD~\cite{SA}. What it says, is that in QCD the
chromomagnetic fields themselves are in general unstable against 
 gluon emission. Although this phenomenon is interesting by itself,
it clearly  has nothing to do with jet energy losses. 
%

The technique developed by Schwinger
\cite{S72} can now be applied to (\ref{SR3}) to derive the power
radiated in a QCD synchrotron process whereby an energetic quark
radiates through a chromomagnetic field. To avoid the unnecessary
algebra triggered by the spin content of the quark, we present
the results for the spin-0 case instead.

For a scalar quark in the fundamental representation, the analogue
of (\ref{SR3}) is

\be
&&{\bf M}_{aa} =
ig^2\,(T^A)_{ab}\,(T^A)_{ba}\,\int\,\frac{d^4k}{(2\pi)^4}\,
\int_{0-i0}^{\infty-i0}\,\frac{ds}{{\rm cos}(g_A\,Hs)}\nonumber\\
&&\times\,
e^{-is(k^2-k_\perp^2({{\rm tan}(g_A Hs)}/({g_A\,Hs}) -1))}
\,(e^{2g_A\,sG})_{\mu\nu}\,\,\Phi (g_A)\nonumber\\
&&\times\,\left(\stackrel{\rightarrow}{\Pi}_a-\stackrel{\leftarrow}\Pi_b\right)^{\mu}\,
((\Pi_b-k)^2-m^2)^{-1} 
\left(\stackrel{\rightarrow}{\Pi}_a-\stackrel{\leftarrow}\Pi_b\right)^\nu\,\,,\nonumber\\
\label{SR7}
\ee
modulo counter-terms. The arrows on $\Pi$'s indicate the 
direction of the derivative. On mass-shell we expect
$\Pi_a\sim \Pi_b+k$, this will hold in the classical
limit.

\subsection{Power Radiated}

For spin-0, the power radiated follows from (\ref{SR8}).
Following Schwinger~\cite{S72} we obtain the chromomagnetic 
synchrotron emission by a scalar quark in the classical limit 
in the following form

\be
{\bf P}_{aa} (\omega) =&&-
\frac {\alpha}{\pi}\,(T^A)_{ab}\,(T^A)_{ba}\,\nonumber\\
&&\times\omega\,{\rm Im}\,\int_{0-i0}^{\infty-i0}\,\frac{d\tau}{\tau}\,
\frac{e^{-i\,(E\omega_A)^2\,{\tau^3}/({24\omega})}}
{{\rm cos}\,({E\omega_A\tau}/({2\omega})}\nonumber\\
&&\times(\frac {m^2}{E^2} + \frac 12\,\omega_b^2\tau^2)\,
e^{-i\omega (\frac {m^2\tau}{2E^2} +\frac {\omega_b^2\tau^3}{24})}
\label{SR9}
\ee
where in (\ref{SR9}) the $H=0$ subtraction is not explicitly shown 
but implied. The quark synchrotron and gluon rescaled frequencies are
$\omega_a=e_a\,H/E$ and $\omega_A=g_A\,H/E$ respectively.

In carrying out (\ref{SR9}) the emitted gluon recoil effect on
the jet was ignored, and so the result
is entirely classical. We have checked that the gluon recoil
effect amounts in the first order to the shift 

\be
\frac 1{\omega}\rightarrow \left(\frac 1{\omega}-\frac1{E}\right)
\label{SR9x}
\ee
in the combination ${\bf P}_{aa}/\omega$  (the gluon multiplicity),
thereby generalizing Schwinger's first quantum correction in QED to 
the QCD case. Of course, this substitution is not the complete quantum 
answer, but will be discussed below as an approximation.

Before we discuss the complete results, let us comment the integrand of
(\ref{SR9}). The last exponent is due to charge curving, and is the
same as in QED. It provides a rapidly oscillating phase at large $\omega$ 
and a corresponding cutoff. The first exponent in the integrand of 
(\ref{SR9}) is new. It stems from the gluon rotation  in the chromomagnetic 
field. The phase follows from the transverse contribution of the gluon 
propagator as is evident from (\ref{SR7}). The cosine in the denominator 
exhibits poles for

\be
\tau_n= \frac {\pi\,\omega}{E\,\omega_A}\,(2n+1)
\label{SR10}
\ee
which are the gluon synchrotron orbits
(classically the gluon spin and the tachyon problem drop). 
The second contribution in (\ref{SR9}) 
is the quark synchrotron contribution as in QED
\cite{LL}.

Rewriting $1/{\rm cos}\,A$ as a geometrical
sum of all powers of $e^{iA}$, we may bring (\ref{SR9}) in the form
of a sum

\be
&&\omega^{-1}{\bf P}_{aa} (\omega) =
\frac {\alpha}{\pi}\,(T^A)_{ab}\,(T^A)_{ba}\,\frac{2m^2}{\sqrt{3}\,E^2}\nonumber\\
&&\times \sum_{n=0}^\infty\,e^{-i\pi\,n\,(1-i0)}\,
\left( {\cal F}(\xi_n) +
2\kappa_n\,{\bf K}_{2/3} (\xi_n)\right)
\label{SR11}
\ee
with

\be
&& {\cal F}(\xi)\equiv \int_{\xi}^{\infty}\,dt\,{\bf K}_{5/3} (t)\,\,,\nonumber\\
&&\left(\frac{\xi_n}{\xi}\right)^2 = \frac{(1+(2n+1)
\,(2\beta)/(3\xi^{2/3}))^3}{(1+\lambda^2)}\,\,,\nonumber\\
&&\kappa_n+1=
\frac{1+(2n+1)
\,(2\beta)/(3\xi^{2/3})}{(1+\lambda^2)}\,\,,
\ee
and

\be
&&\frac{\beta}{\lambda} =
\left(\frac{3\omega_b}{2\omega}\right)^{1/3}\nonumber\\
&&\lambda=\frac{E\omega_A}{\omega\omega_b}\nonumber\\
&&\xi=\frac{2\omega}{3\omega_b}\,\left(\frac{m}{E}\right)^3\,\,.
\label{SR12}
\ee
Again the quantum corrections follow from (\ref{SR11}) through the
substitution (\ref{SR10}) on the RHS. The ${\bf K}$'s are modified
Bessel functions. The sum over $n$ in (\ref{SR11}) sums over synchrotron
orbits of width $-i0$ except for the lowest orbit which is zero.  It is
reminiscent of the sum over `Landau levels' in the Schroedinger formulation.

\subsection{Small and Large $\omega$}

The preceding results are easily analyzed for large and small frequencies
$\omega$. Since the abelian analysis with $g_A=0$ is known from QED, we
focus on the non-abelian part  with $g_A\neq 0$. We will show that the
non-abelian contribution to the synchrotron radiation is strongly
suppressed at small $\omega$ due to strong `incoherence' effects
produced by the deflected radiation. At large $\omega$ the non-abelian
contribution is equal to the abelian contribution. Specifically, the
$g_A\neq 0$ contribution for small $\omega$ reads

\be
&&{\bf P}_{aa} (\omega) \approx \frac {\alpha}\pi \,(T^A)_{ab}\,(T^A)_{ba}\,
\left(\frac{4\sqrt{\pi}}{9\sqrt{\omega}}\right)
\,(E\omega_A)^{3/4}\,\left(\frac mE\right)^2\nonumber\\
&&\times\sum_{n=0}^\infty\,(-1)^n\,(2n+1)^{9/4}\,e^{-\frac 23
(2n+1)^{3/2}\sqrt{E\omega_A}/\omega}\,\,,
\label{AS1}
\ee
which is characterized by an essential singularity at $\omega=0$. 
The $g_A\neq 0$ contribution for large $\omega$ reads

\be
&&{\bf P}_{aa} (\omega) \approx \frac {\alpha}\pi \,(T^A)_{ab}\,(T^A)_{ba}\,
\nonumber\\
&&\times \left(\frac{2\sqrt{\pi}}{9}\right)
\,\left(\frac {m\omega}{E\omega_b}\right)^{5/2}
\omega_b\, e^{-(m/E)^3\,(2\omega/3\omega_b)}\,\,.
\label{AS2}
\ee

At small $\omega$ the non-abelian radiation is of order
$e^{-\#/\omega}/\sqrt{\omega}$ which is much smaller than the
abelian radiation of order $\omega^{1/3}$. The reason is that the 
emitted non-abelian gluon brings about its own phase which strongly
adds to the phase incoherence at small $\omega$ as is clearly seen in
(\ref{SR3}) and (\ref{SR7}). At large $\omega$ both the abelian and
non-abelian radiations are comparable and of order
$\omega^{5/2}\,e^{-2\omega/(3\omega_b\gamma^3)}$. Indeed, we note that
(\ref{AS2}) is independent of the gluon charge $g_A$. For
ultrarelativistic jets, the radiation frequencies are in the
range $0\leq \omega\leq E$. Thus $\omega/\omega_b\leq \gamma$ which
is way below the maximum of $\gamma^3$.
In light of the present observations we conclude that most of the jet
radiation is emitted through the `abelian' part of the gluon charge
and in the small frequency range way below the synchrotron maximum since
$E/\omega_b \ll \gamma^3\,$. The `non-abelian' part only enters as a
correction.

\subsection{Energy Loss}

\begin{figure}[h]
\epsfxsize=6.cm
\centerline{\epsffile{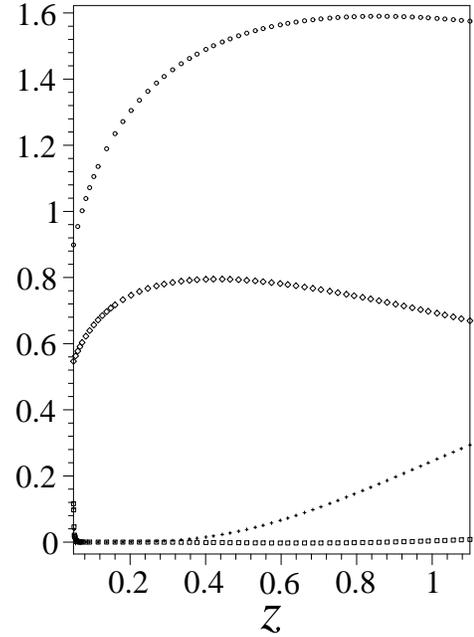}}
\centerline{\epsffile{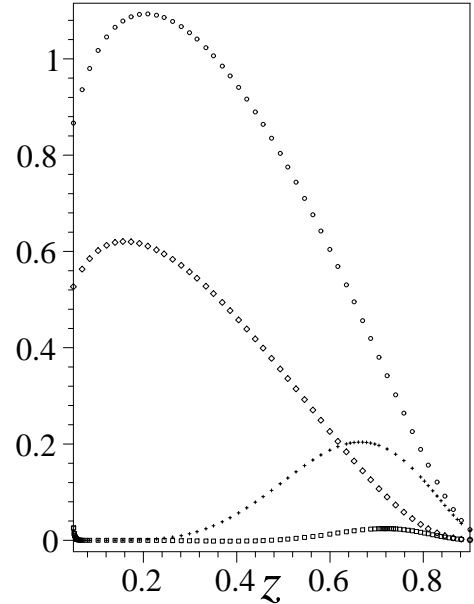}}
\vskip 0.05in
\caption[]{
 \label{fig_quantumcut2}
The integral in (\ref{eqn_int_I}) versus the 
radiated energy fraction $z=\omega/E$ 
for classical (upper figure) and quantum corrected 
(lower figure) cases. Four examples are shown
corresponding to all possible charge combinations, namely
$ e_a=1,e_b=1,g_A=3$ (boxes), $e_a=1,e_b=-2,g_A=3$
(crosses), $e_a=1,e_b=1,g_A=0$ (diamonds) and 
$e_a=1,e_b=-2,g_A=0$ (circles).
}
\end{figure}

The `abelian' part of the jet radiation follows from the summation
over $A=1,2,3,8$ for which $g_A=0$, thus $e_a=e_b$. In this case
(\ref{SR11}) simplifies to the QED-like answer

\be
\omega^{-1}{\bf P}_{aa} (\omega) =&&
\frac
{\alpha}{\pi}\,(T^A)_{ab}\,(T^A)_{ba}\,\frac{m^2}{\sqrt{3}\,E^2}{\cal F}(\xi)
\label{SR13}
\ee
since $\xi_n=\xi$ and $\kappa_n=0$. In this case, the total power
emitted follows by integrating (\ref{SR13}) over the gluon frequency,
explicitly including the recoil effects (\ref{SR11}),

\be
{\bf P}_{aa}
=&&\frac{\alpha}{\pi}\,(T^A)_{aa}\,(T^A)_{aa}\,\frac{m^2}{\sqrt{3}}
\nonumber\\&&\times \int_0^E\,
\frac{d\omega}{E}\,\frac{\omega}{E}\,
{\cal F}(\xi_{\rm corr})
\label{SR14}
\ee
with a different (quantum corrected)
$\xi_{\rm corr}$

\be
\xi_{\rm corr}=\frac
{2}{3}\,\left(\frac{m}{E}\right)^3\,
\frac {\omega/\omega_a}{1-{\omega}/E}\,\,.
\label{SR15}
\ee
In Fig.~\ref{fig_quantumcut2} we compare two spectra for some particular 
selection of parameters.
The general result for (\ref{SR14}) can be obtained by expanding around
the classical result with the first quantum correction included


\be
{\bf P}_{aa}
\approx \frac{\alpha}{\pi}\,(T^A)_{aa}\,(T^A)_{aa}\,\,{\bf
C}\,\,(\omega_a\,E^2)^{2/3}
\label{SR15x}
\ee
with the constant

\be
{\bf C} =\frac {-\pi\,(3/2)^{8/3}}{10\,{\rm sin}\,(\pi/3)}\,\frac{2^{5/3}}{\Gamma(-2/3)}\,
\,(1-\frac 8{21})\approx 0.52\,\,.
\ee
The contributions in $(1-8/21)$ are the classical contribution 
and the first quantum recoil correction respectively. So the 
recoil of the emitted gluon decreases the net radiation by a factor 0.62. 
The ratio of the contributions of the classical to recoil corrected 
spectra  is 0.48. Within few percents these results agree with direct
numerical estimates of the integrals including the approximate all orders
quantum recoil corrections. This justifies a posteriori the use of the
first order correction in our calculations.

In order to further compare losses related with emission of QED-like and
QCD-like gluons, we show in Fig.~\ref{fig_quantumcut2} the leading (n=0) 
integral

\be \label{eqn_int_I}
&&{\bf I}=\int_{0}^{\infty}\,\frac{d\tau}{\tau}\,(\frac {m^2}{E^2} + \frac
12\,\omega_b^2\tau^2)  \nonumber\\
&&\times {\rm sin}\,
[{(E\omega_A)^2 \,\tau^3 \over 24\omega}+{E\omega_A\tau \over 2\omega}
+  {\omega m^2\tau \over 2E^2} + {\omega \omega_b^2\tau^3 \over 24}] 
\ee 
with the $H=0$ subtraction implied,
for $E=m=H=1$ and 4 different sets of charges (see the figure caption).

%
%

\section{Applications to heavy ion collisions}

In this section we will first recall for completeness
some qualitative arguments regarding jet quenching in 
heavy ion collisions, most of which have already been 
discussed in~\cite{Shu_how}. We will then use the above
synchrotron radiation results to assess jet energy loss
in the color glass condensate approach and the exploding 
sphaleron approach.

\subsection{Introduction to Jet Quenching}

Jet quenching is  a sort of ``tomography''of the prompt
excited system, created in high energy heavy ion collisions. 
Even very hard jets radiate and lose some energy
during their passage through the system, thereby providing
information about the early stages of the collision.
 
The so called {\em quenching factor} $Q(p_t)$  is defined as
the observed number of jets normalized to the $expected$ number 
of jets calculated in the parton model {\em without} account 
for final state interactions~\footnote{
Effects due to $initial$ state interaction should be
also included. What this means is that the parton distribution 
functions should be nuclear rather than hadronic, following from
lepton-nuclei experiments. Parton rescattering in nuclei at 
the origin of the so called Cronin effect, should also be included in
the expected yield. }. It is usually assumed that hard QCD probes 
are under good theoretical control, and that one can assess 
the initial jet production reliably.

Experimentally, jet reconstruction in a
heavy ion environment is very difficult to achieve. Therefore,
all currently reported results  for jet quenching refer to the 
observed/expected ratio of the yields of  {\em single hadrons}. 
Furthermore and in so far,  large $p_t$ means $p_t =$2-6 GeV.
Which part of this transverse momentum comes from genuine high
energy jets is anybody guess. The first direct evidences for jets have
been recently reported by the STAR collaboration~\cite{STAR_jet},
whereby a second particle correlated with the trigger  with 
$p_t> 4$ GeV was observed, well inside a relatively narrow cone 
typical of jets.

In the early theoretical studies on the subject~\cite{early},
a rather modest jet re-scattering in the 
Quark-Gluon Plasma (QGP) has been considered.
Accounting for the radiation effect~\cite{Gyulassy_losses} 
has significantly increased expectations for the
magnitude of the result, while accounting for 
the Landau-Pomeranchuck-Migdal (LPM) effect~\cite{Dok_etal} 
has somewhat decreased the magnitude of the result. We will not
discuss this involved subject, but only recall
that the expected quenching factor from such studies is 
$Q(p_t)=$0.5-0.7 for jets with $p_t=$10-20 GeV.

Experimentally, a relatively modest jet quenching has been first
observed in deep inelastic scattering for a forward jet
going through cold nuclear matter (for recent discussion and references
see~\cite{WangWang}). The heavy ion data at the CERN SPS
have also shown modest quenching effects, but
already the very first RHIC data \cite{QM01}
(especially for $\pi^0$ from the PHENIX collaboration) have
shown that quenching of jets  is very strong, with $Q (p_t) <
1/3$. Subsequent discoveries that (i) at 
$p_t> 2$ GeV protons and anti-protons
dominate the charge particle spectra; 
(ii) that the azimuthal asymmetry
remains very strong  even at large $p_t$; (iii) that development of very
successful hydro and/or cascade description of spectra without jets  
even at  $p_t=$2-3 GeV,  have all led to suspect 
that the real jet quenching factor maybe even stronger.  
Furthermore, one of us  even found \cite{Sh_v2_largept} 
that the very high degree of azimuthal asymmetry  observed by
the STAR experiment at $p_t=$2-6 GeV~\cite{STAR_jet} cannot be
reproduced by {\it any} amount of jet quenching, no matter how
strong.

One technical but important point made in the last paper
in~\cite{Dok_etal}, is that when the  quenching is strong it cannot be
evaluated using the {\it mean} energy loss. Specifically,
the quenching factor can be seen as the ratio of produced-and-quenched 
to produced spectra

\be 
Q(p_t)={\int d\epsilon D(\epsilon)dN/d
p_t^2(p_t+\epsilon) \over  dN/d
p_t^2(p_t)}
\label{one}
\ee	
where $\epsilon$ is the energy lost in the medium and $ D(\epsilon)$
its normalized distribution. For small $\epsilon$ one can
expand it to first order, obtaining a correction proportional to
the mean energy loss $<\epsilon>= \int d\epsilon D(\epsilon)\epsilon$.
However, because the spectrum is so steep, this is only valid when
$\epsilon/p_t$ is not larger than few percent.

Therefore and for a qualitative assessment of the magnitude of the
effect needed, we suggest a different
simple approximation. Using a power parameterization of the spectrum

\be 
{dN \over d p_t^2(p_t)} \sim {1 \over p_t^n}
\ee
for both the observed and ``hard'' distributions, we obtain

\be  \label{quenching_factor}
Q(p_t)=\int d\epsilon D(\epsilon)({1 \over 1+\epsilon/p_t})^n
\sim \int d\epsilon D(\epsilon)e^{-n\epsilon/p_t} 
\ee
instead of (\ref{one}). Using a simple delta-like distribution peaked
at some fractional loss,

\be 
D(\epsilon)=\delta(\epsilon-\kappa p_t) 
\ee
we have $Q(p_t)=1/( 1+\kappa)^n$. With 
$n\approx 12$ in the few-GeV domain at RHIC energies,
a jet quenching by one order of magnitude would correspond to
$\kappa\sim 1/4$. This means that  a mean loss of about 
15-20\% of the produced jet momentum is sufficient. 
However, this conclusion is oversimplified. As
one can see from (\ref{quenching_factor})
the quenching factor is dominated by small losses $\epsilon/p_t<1/n\sim
1/12$. What this means is that what matters is the probability to escape 
with as small losses as possible. 

This conclusion changes the relative role of early (synchrotron-like)
versus late (multiple bremsstrahlung with LPM) effects. While the
latter can be very large for specific geometry 
(LPM energy loss \cite{Dok_etal} is $\Delta E \sim
L^2$ where L is the path in matter), the integral
(\ref{quenching_factor}) would  be dominated by surface emission with
small $L\sim 1$ fm or so. Early effects emphasized in this paper,
even if producing less average losses, are expected to affect
the  probability $D(\epsilon)$ at small $\epsilon$, preventing 
easy escape of some jets. Detailed numerical simulations (which are
well beyond the limits of this work) are needed to understand their
relative role.

Summarizing this subsection, we say that the
traditional approach in which excited matter 
such as a QGP consists of a collection of 
{\rm uncorrelated} quarks an gluons acting as
scattering centers for the high energy partons,
has difficulties accounting for the large jet
losses reported at RHIC. This is the primary motivation
for considering synchrotron-like radiation in this work.

\begin{figure}[h]%
\leavevmode
 \begin{minipage}[c]{4.cm}
 \centering 
\includegraphics[width=3.cm]{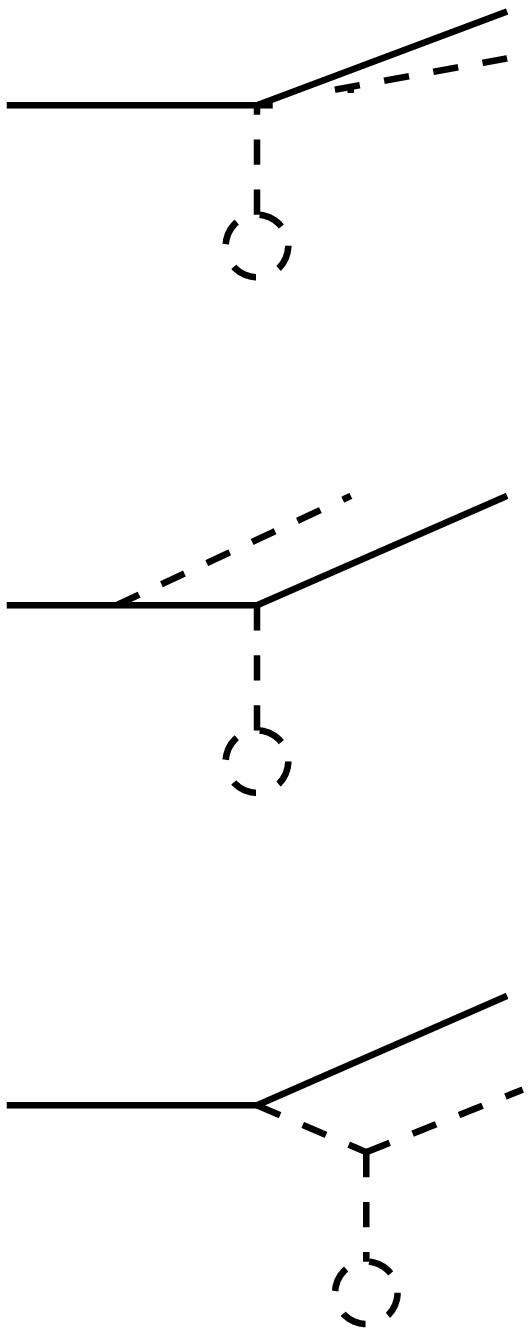}
 \end{minipage}
   \begin{minipage}[c]{4.cm}
\centering 
\includegraphics[width=3.cm]{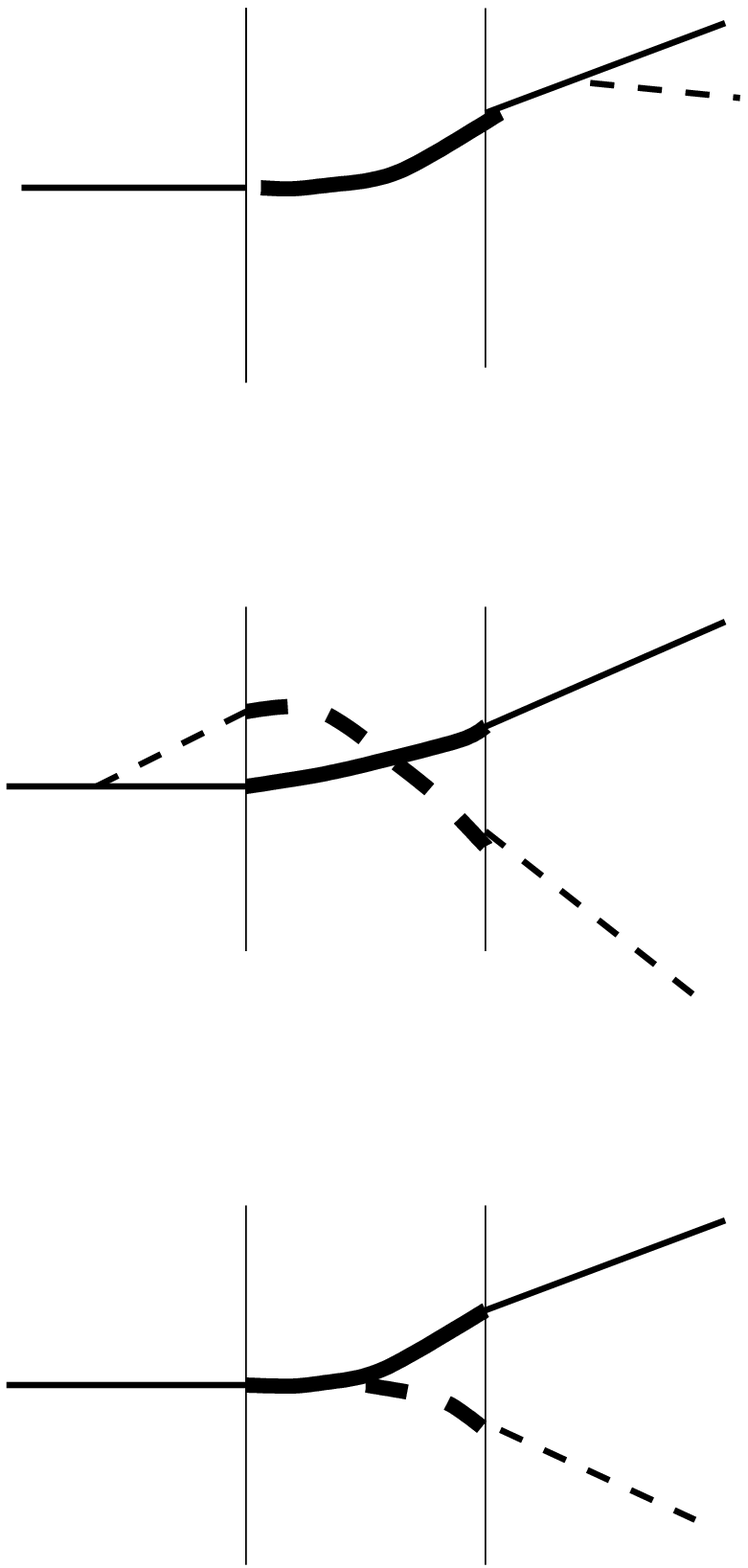}
\end{minipage}
\caption[]{
 \label{fig_BG}
The l.h.s. shows three diagrams illustrating QCD bremsstrahlung
radiation: the small  circle represents the source of the perturbative 
field. The r.h.s. shows three 
similar diagrams illustrating  QCD bremsstrahlung radiation
(a,b) and  synchrotron-like radiation (c). The strong chromomagnetic field
is assumed to be in between two thin vertical lines, where the full (dressed) 
propagators are indicated by thicker lines.
}
\end{figure}

\subsection{QCD bremsstrahlung}

For completeness we start with a  brief reminder of  ordinary QCD 
bremsstrahlung. This effect was first considered by Bertsch and 
Gunion~\cite{BG}, who 
calculated the three diagrams shown in the l.h.s of Fig.~\ref{fig_BG} for
soft gluon emission. Adding their squares~\footnote{The interference terms 
were also discussed in that paper. They are related to form
factors which are relevant to radiation inside hadrons, which we are not 
interested in here.} at high collision
energies relative to both the momentum transfer $\vec l_t$ from the target
parton and also the  transverse momentum of the radiated gluon $\vec q_t$, 
they have shown that the number of gluons emitted is

\be \label{eq_GB}
{dN_g \over dy d^2 q_t}= {C^2_c \alpha_s \over \pi^2} {\vec l_t^2
\over
 \vec q_t^2 (\vec q_t-\vec l_t)^2}
\ee 
where $C^2_c$ is a color factor (3 for qq scattering with gluon emission).
Note that at zero momentum transfer $\vec l_t=0$ the radiation 
vanishes. Also, when the denominators become small, they generate
two cones of radiation, along the initial and final direction of the 
jet. Integrating this result over the rapidity $y$ and the transverse momentum 
$q_t$ of the gluon, yields the standard logarithmic factor.

In our case the field crossed by a jet parton
is {\em classical and  non-perturbative} with  $A\sim 1/g$, and
the radiation is described by the modified diagrams shown in the r.h.s. of
Fig.~\ref{fig_BG}. The motion inside the field is described by fully
dressed propagators, and the power radiated from the slab has been described
above. Assuming the chromomagnetic field to be confined to a finite slab, 
requires that we also add the bremsstralunhg diagrams from the in and out
motion as illustrated in the first two diagrams of~Fig.~\ref{fig_BG}. 
The standard QCD variant of the QED Weizsacker-Williams (WW) 
approximation can be used, with  the so called ``parton-in-parton'' DGLAP 
splitting functions~\cite{Shu_how}

\be \label{p-in-p}
 {\partial\phi_{i/j}(x,\mu) \over \partial {\rm ln}\,\mu}=
{\alpha_s(\mu) \over 2\pi} P_{ij}(x)
\ee
where $x$ is the standard parton momentum fraction and i,j=q,g. 
The splitting functions are known for all values of $x$, and the
total energy loss for quark/gluon due to splitting is~\cite{Shu_how}

\be 
{\epsilon_q \over p}\approx
0.28 \,\alpha_s\, {\rm ln}\,({Q^2_{\rm high} / Q^2_{\rm low}} ) 
\ee
\be {\epsilon_g \over p}\approx 1.0\,
\alpha_s\, {\rm ln}\,({Q^2_{\rm high} / Q^2_{\rm low}} ) 
\ee
where the second integral was regulated by setting $x_{\rm max}=m^*/P$, 
taken to be 0.95 in the estimate. The parton-to-parton splitting
happens twice, i.e. in and out. On the way in, the 2 scales that 
define the DGLAP evolution of the jet are the kick in the scattering
process and the parton virtuality while hitting the magnet. On the
way out, the 2 scales are the kick from the slab and the final scale
of the parton in matter or its hadronization scale (whichever is larger).

We close this summary section by the following qualitative
comments: i. The strong bending of the partons happens rarely,
so it is not subject to the LPM effect; ii. The in and out 
bremsstrahlung effect depends weakly (logarithmically) on the
field strength provided it is large enough to allow for the
separation of the 2 cones of radiation; iii. The magnitude
of the relative energy loss by bremsstrahlung alone is of order 
$\Delta E/E\sim 1/4$, and maybe alone sufficient to explain 
the expected energy loss for gluon jets at RHIC; iv. The
contribution from diagram (a) is in general small if the
jet has been just produced in a hard collision, since the
virtual field does not have sufficient time to form.

\subsection{Jet quenching in the Color Glass Condensate}

If the initial excited glue is not a set of incoherent gluons
with occupation numbers $n\sim O(1)$, as it is the case in an
equilibrated QGP, but a coherent classical field with large 
occupation numbers $n\sim O(1/g^2)$,
the  radiation losses are synchrotron-like. Coherence 
helps, because all coherent quanta work in the
same direction in space and color space, providing larger
acceleration. 

To assess the amount of synchrotron radiation in the CGC phase,
we recall some useful numbers from the numerical analysis of the
SU(3) version of the CGC carried in~\cite{KNV}. At RHIC energies
the saturation scale $Q_s$ was found to be 1.3 GeV. The initial
classical CGC field was found to abelienize in a time $\tau_{\rm CGC}$ 
of order 
$Q_s\tau_{\rm CGC}\sim 3$. In this regime, the gluon energy density was found
to be $\epsilon/Q_s^4\approx 0.17/g^2$ with an approximatly
thermal momentum distribution. The transverse energy per quantum 
was found to be $1.66\, Q_s$, resulting in an  effective temperature 
of 1 GeV. This is of course only apparent as the underlying evolution
is classical and originates from a coherent state. The field strength $F$ 
(the r.m.s. combination  of  electric and magnetic fields) is about 

\be 
gF\sim 0.58\, Q_s^2\sim 1\, {\rm GeV}^2 
\ee 
giving a quark with $e_a=g\sqrt{3}/6$ a kick of order 
$e_a F \tau\sim 0.66$ GeV, and about twice that for a gluon.

Before substituting these numbers into our expressions for the
synchrotron radiation loss, we need to do some relevant color sums

\be 
C_A=\sum_a (T^A)^2_{aa}\,
|e_a|^{2/3} 
\ee
The values for ``penetrating gluons'' of kind 3 and 8 are 0.30 and 0.22,
respectively. For all gluons, we have $\sum_{A=1,8} C_A=2.06$, with about 
$1/2$ originating from the undeflected gluons and $1/2$ originating from
the deflected ones (of course with the extra penalty factor at low 
$\omega$ as explained above). For the estimate to follow we will use
a color factor of 1.5.

Using the CGC numbers just quoted,  we find that the relative energy
loss of a quark by synchrotron radiation in a time $\tau_{\rm CGC} $is

\be 
{\Delta E_{\rm CGC} \over E}\approx 0.3 \,
\left({H  \over  1\, {\rm GeV}^2}\right)^{2/3}
\left({ \Delta \tau_{\rm CGC} \over 0.5 \, fm}\right)
\left({ 1 \, {\rm GeV} \over E}\right)^{1/3}\,\,.
\ee
The gluon loss is about twice the quark loss.

\subsection{Jet Quenching on the Exploding Sphalerons}
\label{sec_sphal}

The energy and the Chern-Simons number of the released (turning) 
coherent state is

\be
E=&&\frac {3\pi}{4\alpha\rho}\,{\cal E}\nonumber\\
N_{\rm CS} = &&\frac 12\, {\cal N}
\ee
In so far, the dimensionless parameters have been determined 
in two ways. First, by minimizing the QCD potential  
for fixed Chern-Simons number with the result~\cite{OCS}

\be
{\cal E} = &&(1-\kappa^2)^2\nonumber\\
{\cal N} = &&{\rm sign}(\kappa)(1-|\kappa|)^2(2+|\kappa|)/2
\ee
Eliminating $\kappa$ yields the potential profile $E(N_{\rm CS})$.
Second, by maximizing the partial parton-parton cross section
with the result~\cite{JSZ}

\be
{\cal E} = &&(E/M_S)\nonumber\\
{\cal N} = &&(E/M_S)^{2/5}
\ee
At the sphaleron point the partial cross section is maximum,
with ${\cal E}={\cal N}=1$. Only this case will be considered
here. Using instanton vacuum physics we obtain a sphaleron mass
$M_S=3$ GeV and a sphaleron size $\rho=1/3$ fm.

Upon release in Minkowski space, the sphaleron state evolves 
classically in real time through the classical Yang-Mills
equations, as was originally done in the electroweak theory
~\cite{sphaleron_decay}. For SU(2) Yang-Mills, this evolution
was recently carried out in~\cite{OCS} both numerically and
analytically. The turning states were found to explode into
thin shells of coherent gluonic fields. For our purposes, we
just recall that the shell for $t,r\gg \rho$ has a very simple
radial energy density

\be \label{eqn_wall}
4\pi r^2 e(r,t) = \frac{8\pi}{g^2\rho^2}\,
\left(\frac{\rho^2}{\rho^2+(r-t)^2}\right)^3
\ee
At large time $t\gg \rho$ the corresponding gauge field is purely 
transverse, with equal chromoelectric and chromomagnetic fields.
The prompt sphaleron configuration released in parton-parton scattering
carries initially a very strong chromomagnetic field, 

\be
\sqrt{H}\sim \left(\frac{2M_S}{\rho^3}\right)^{1/4}\sim 1\,\,{\rm GeV}\,\,.
\label{1}
\ee

In the early phase of the prompt process in heavy ion collisions,
the escaping sphalerons form a dilute gas. So unlike the CGC they 
cannot affect most of the jets initially for times $t\sim \rho$.
They do affect them as they expand into exploding shells. The net 
synchrotron radiation loss involves also the transverse
density of sphalerons per unit rapidity $n_S$ and their typical collision 
volume $\sigma (t)\, dt$. Specifically,

\be 
\Delta E=\int\, {\bf P}(t)\, n_{S}\,\sigma(t) dt\sim
\int\, dt\, t^{-2/3+2} 
\ee
where we used the cross section $\sigma\sim t^2$ and the radiation loss
${\bf P}\sim H^{2/3}\sim t^{-2/3}$. The result formally diverges for
large times. However, the above reasoning is only valid till the single
shell expansion remains coherent. As we now show, the originally dilute
gas of shells quickly evolves into a foam-like structure for times 
$t\sim$ 2-3$\rho$ providing a natural cutoff in the time integral.

Recent estimates of the number of clusters produced in pp and AA
collisions as a function of the collision energy and centrality,
are still rather uncertain. The theoretical calculations of the
cross section such as~\cite{JSZ} are carried to only exponential
accuracy, while phenomenological studies such as~\cite{COS1} have
only resulted into an upper estimate. Assuming the whole growth 
with $\sqrt{s}$ of the pp cross section to be due to the release of
1 sphaleron, and ignoring nuclear shadowing, about ${\bf S}=400$ sphalerons 
are produced in central AuAu collision at $\sqrt{s}=136$ GeV at RHIC.
Since the total rapidity interval is $\Delta y=4$, this amounts to
about 100 sphalerons per unit rapidity. The transverse density 
of sphalerons per unit rapidity in central AuAu collisions is

\be 
n_{S}={{\bf S} \over \pi R^2 \Delta y }\sim 1\, {\rm fm}^{-2}
 \ee
which results into a foam-like structure for the exploding 
sphalerons for times $t\sim 2\rho\sim 0.7$ fm. This time
increases as the cubic root of the transverse density for 
smaller densities. For a jet piercing a wall with the 
energy density (\ref{eqn_wall}) the integrated  kick is about
0.5 GeV for a quark with $e_a=g\sqrt{3}/6$, and about 1 GeV 
for a gluon.

Substituting the expression for the
field strength of the exploding shell (\ref{eqn_wall}) into 
the synchrotron radiation loss  (\ref{SR15x}), yields 
$ \int F^{2/3} dt=F^{2/3}_{\rm max}\pi\rho$. For two overlapping shells we add
the fields in quadrature and estimate that the maximal field $F_{\rm max}\approx 0.2 \,
{\rm GeV}^2$ in the foam phase at time $t_{\rm foam}\approx 2 \rho$. 
Substituting all this into the synchrotron energy loss formulae, we finally 
obtain  for the quark loss

\be 
{\Delta E \over E}\sim 0.21\, 
\left({H  \over  0.2\, {\rm GeV}^2}\right)^{2/3}
\left({ 1 \, {\rm GeV} \over E}\right)^{1/3}\,\,.
\ee
The gluon loss scales with the pertinent color Casimir and is about 
twice larger.

\section{Summary and discussion}

The chief idea of this work is that if the initial stage of a heavy ion 
collision produces a coherent classical field rather than an a gas of
incoherent quanta, one should reconsider the theory of {\it all} 
prompt processes (Drell-Yann, photons, dileptons, heavy quarks, ...),
including the current theory of jet energy losses. Instead of multiple 
small angle scattering subject to the Landau-Pomeranchuck-Migdal suppression, 
we have synchrotron-like QCD radiation. The radiation is enhanced by the
coherent  classical fields, providing larger acceleration and radiation
compared to  independent quanta.

There are currently two mechanisms for the formation of strong and prompt
classical color fields in heavy ion collisions. First, the color
glass condensate (CGC)~\cite{McLV} is a classical Weiszacker-Williams 
field of virtual gluons initially part of the wavefunction of the
colliding nuclei, that is made real by the collision. Second, the
exploding sphaleron-like clusters which are the remnants of (singular) 
instantons in the QCD vacuum. The clusters are not only coherent,
but evolve into thin shells of strongly localized fields and become
foam-like. Any jet has a pobability of about 1 to interact with
walls of exploding clusters.

We have shown that synchrotron radiation loss from a 6 GeV quark jet 
is about 0.17 in the CGC and about 0.12 in the exploding sphalerons.
Gluon jet losses are about twice larger. The corresponding bremsstrahlung 
radiation loss on the in and out motion results also in a loss of a similar
magnitude. At larger jet energies, bremsstrahlung loss dominates over 
synchrotron loss, although the latter is found to decrease slowly
with increasing energy (as $1/E^{1/3}$). All in all, the mechanisms
of jet quenching considered here provide in total about 20-30\% loss,
which is about consistent with the empirically reported jet quenching at
RHIC.

The present considerations of jet quenching in the context of
the color glass condensate or the exploding sphalerons is rather
schematic, with quantum effects carried only to leading order. 
As emphasized above, however, what really matters is not the
average radiation loss for a jet, but rather the probability for
a jet to escape without losses. To assess this, detailed simulations
with realistic nuclei geometry are needed. Also, as suggested
in~\cite{Shu_how}, it is possible to experimentally measure whether
the radiated gluons are produced in a narrow cone around the jet
(bremmstrahlung) or not (synchrotron). One can also do tagged jets
by measuring photon-jet correlations, and search for acomplementarity 
and $p_t$ disbalance. Clearly much more theoretical work is needed to 
make the present estimates more quantitative.

\begin{acknowledgments}
This work was  partially supported by the US-DOE grants DE-FG02-88ER40388
and DE-FG03-97ER4014.
\end{acknowledgments}

\appendix

\section{Classical reduction}

In this Appendix we give some helpful steps leading to the
classical formulae (\ref{SR9}). We will show how to reduce the
gluon part, since the quark part follows from Schwinger's
paper~\cite{S72} to which we refer to. We first exponentiate the 
scalar quark propagator in (\ref{SR7}), combine it with
the already exponentiated gluon propagator and change 
the proper-time variables (Feynman parametrization) to obtain 

\be
&&\int_0^\infty\,s\,ds\,\int_0^1\,
\frac{du}{{\rm cos}(g_A\,Hs(1-u))}\,e^{-ism^2u}\,e^{-is\,{\bf
H}_b}\nonumber\\
&&\times e^{-is (1-u)\,k_\perp^2(({{\rm tan}(g_A Hs(1-u))}/({g_A\,Hs(1-u)}) -1)}\,\,,
\label{A1}
\ee
for the propagators only. The proper-time hamiltonian is

\be
{\bf H}_b = (k-u\Pi_b)^2 + u(1-u)\,\Pi_b^2\,\,,
\label{A2}
\ee
with a `mass-shell' condition $\partial{\bf H}_b/\partial k\sim 0$
leading to $k\sim u\Pi_b$ or $k^2\sim u^2m^2$ in the classical
limit. This saddle point relation receives corrections at the
quantum level.

Following~\cite{S72} we introduce the key change of
variables that facilitates the identification with the classical
radiation problem~\cite{LL}

\be
&&s=\frac{\tau}{2\omega}\nonumber\\
&&u=\frac{\omega}E\ll 1\,\,,
\label{A3}
\ee
where the last inequality will be relaxed through the first quantum
correction.
Using (\ref{A3}) and the substitution $ k_\perp^2\rightarrow \omega^2$
valid to leading order in $\omega/E$ (classical), we find the classical
limit to the gluonic contribution to be

\be
\frac{e^{-i\,(E\omega_A)^2\,{\tau^3}/({24\omega})}}
{{\rm cos}\,({E\omega_A\tau}/({2\omega})}
\label{A4}
\ee
which is the part quoted in (\ref{SR9}). The remaining quark part
follows exactly Schwinger's argument and will not be repeated here.

\end{document}